\documentclass[12pt]{article}

\usepackage[a4paper,total={18cm,27cm}]{geometry}
\usepackage{graphicx}
\usepackage{authblk}
\usepackage{amsmath}
\usepackage[utf8]{inputenc}
\usepackage{hyperref}
\usepackage{verbatim}
\usepackage{color,soul}
\usepackage{caption}
\usepackage{xcolor}
\usepackage{framed}
\usepackage{array}
\colorlet{shadecolor}{yellow!20}

\newcommand{\NI}{\vspace{0.2cm}\noindent}

\begin{document}




\title{Recurrence Resonance - \\ Noise-Enhanced Dynamics in Recurrent Neural Networks}


\author[1]{Claus Metzner}
\author[1,2]{Achim Schilling}
\author[1]{Andreas Maier}
\author[1,2]{Patrick Krauss}

\affil[1]{\small Cognitive Computational Neuroscience Group, Pattern Recognition Lab, Friedrich-Alexander-University Erlangen-Nürnberg (FAU), Germany}
\affil[2]{\small Neuroscience Lab, University Hospital Erlangen, Germany}

\maketitle


\vspace{0.4cm}
\begin{abstract}
\NI Understanding how neural networks process information is a fundamental challenge in neuroscience and artificial intelligence. A pivotal question in this context is how external stimuli, particularly noise, influence the dynamics and information flow within these networks. Traditionally, noise is perceived as a hindrance to information processing, introducing randomness and diminishing the fidelity of neural signals. However, distinguishing noise from structured input uncovers a paradoxical insight: under specific conditions, noise can actually enhance information processing. This intriguing possibility prompts a deeper investigation into the nuanced role of noise within neural networks. In specific motifs of three recurrently connected neurons with probabilistic response, the spontaneous information flux, defined as the mutual information between subsequent states, has been shown to increase by adding ongoing white noise of some optimal strength to each of the neurons \cite{krauss2019recurrence}. However, the precise conditions for and mechanisms of this phenomenon called 'recurrence resonance' (RR) remain largely unexplored. Using Boltzmann machines of different sizes and with various types of weight matrices, we show that RR can generally occur when a system has multiple dynamical attractors, but is trapped in one or a few of them. In probabilistic networks, the phenomenon is bound to a suitable observation time scale, as the system could autonomously access its entire attractor landscape even without the help of external noise, given enough time. Yet, even in large systems, where time scales for observing RR in the full network become too long, the resonance can still be detected in small subsets of neurons. Finally, we show that short noise pulses can be used to transfer recurrent neural networks, both probabilistic and deterministic, between their dynamical attractors. Our results are relevant to the fields of reservoir computing and neuroscience, where controlled noise may turn out a key factor for efficient information processing leading to more robust and adaptable systems.
\end{abstract}

\section{Introduction}


\NI Artificial neural networks are a cornerstone of many contemporary machine learning methods, especially in deep learning \cite{lecun2015deep}. Over the past decades, these systems have found extensive applications in both industrial and scientific domains \cite{alzubaidi2021review}. Typically, neural networks in machine learning are organized in layered structures, where information flows unidirectionally from the input layer to the output layer. In contrast, Recurrent Neural Networks (RNNs) incorporate feedback loops within their neuronal connections, allowing information to continuously circulate within the system \cite{maheswaranathan2019universality}. Consequently, RNNs function as autonomous dynamical systems with ongoing neural activity even in the absence of external input, and they are recognized as 'universal approximators' \cite{schafer2006recurrent}. These unique characteristics have spurred a significant increase in research on artificial RNNs, leading to both advancements and intriguing unresolved issues: Thanks to their recurrent connectivity, RNNs are particularly well-suited for processing time series data \cite{jaeger2001echo} and for storing sequential inputs over time \cite{schuecker2018optimal,busing2010connectivity,dambre2012information,wallace2013randomly,gonon2021fading}. For example, RNNs have been shown to learn robust representations by dynamically balancing compression and expansion \cite{farrell2022gradient}. Specifically, a dynamic state known as the 'edge of chaos', situated at the transition between periodic and chaotic behavior \cite{kadmon2015transition}, has been extensively investigated and identified as crucial for computation \cite{wang2011fisher,boedecker2012information,langton1990computation,natschlager2005edge,legenstein2007edge,bertschinger2004real,schrauwen2009computational,toyoizumi2011beyond,kaneko1994evolution,sole1995information} and short-term memory \cite{haruna2019optimal,ichikawa2021short}. Moreover, several studies focus on controlling the dynamics of RNNs \cite{rajan2010stimulus,jaeger2014controlling,haviv2019understanding}, particularly through the influence of external or internal noise \cite{molgedey1992suppressing, ikemoto2018noise,krauss2019recurrence,bonsel2021control,metzner2022dynamics}. RNNs are also proposed as versatile tools in neuroscience research \cite{barak2017recurrent}. Notably, very sparse RNNs, similar to those found in the human brain \cite{song2005highly}, exhibit remarkable properties such as superior information storage capacities \cite{brunel2016cortical} \cite{narang2017exploring,gerum2020sparsity,folli2018effect}.

\NI In our previous research, we systematically analyzed the relation between network structure and dynamical properties in recurrent three-neuron motifs \cite{krauss2019analysis}. We also demonstrated how statistical parameters of the weight matrix can be used to control the dynamics in large RNNs \cite{krauss2019weight, metzner2022dynamics}. Another focus of our research are noise-induced resonance phenomena \cite{bonsel2021control, schilling2022intrinsic, krauss2016stochastic, schilling2021stochastic, schilling2023predictive}. In particular, we discovered that in specific recurrent motifs of three probabilistic neurons, connected with ternary ($-1,0,+1$) weights, the mutual information $I$ between subsequent system states can go through a resonance-like maximum when normal-distributed white noise of increasing standard deviation $r$ is added independently to all neurons. The phenomenon was called 'Recurrence Resonance' (RR) \cite{krauss2019recurrence}, because $I$ can be interpreted as the spontaneous recurrent information flux in the network. It grows with the number of visited system states and with the degree of predictability of each successor state from its predecessor.

\NI Since $I$ is a key factor for the information processing faculties of RNNs, it became important to understand which types of weight matrices enable a large spontaneous information flux in probabilistic RNNs, such as Symmetric Boltzmann Machines (SBMs, see Methods for details). By reverse-engineering evolutionary optimized networks \cite{gerum2020sparsity}, we found indeed a universal design principle for maximizing $I$ \cite{metzner2024quantifying}. It was called 'NRooks', because in optimal N-neuron networks each row and column of the $N\!\times\!N$ weight matrix only contains a single non-zero entry, resembling the N-rooks-problem in chess \cite{katz1972coverage}. While these $N$ non-zero elements should ideally have the same magnitude $w$, their signs can be arbitrary. In the limit of large magnitudes $w$, the SBMs become quasi-deterministic and the information flux $I$ approaches the theoretical maximum $I_{max}\!=\!N$. In this extreme case, all $2^N$ possible system states are periodically visited in a fixed order.

\NI The present work aims to understand, on a deeper level than before, the pre-conditions of the RR phenomenon, as well as its mechanism. As model systems, we will mainly use probabilistic SBMs, but we will also briefly consider deterministic networks with '$\tanh$' activation functions. Different types of weight matrices will be investigated, but NRooks system will play a particularly important role, as their information theoretic properties are very well understood \cite{metzner2024quantifying}.

\section{Methods}

\subsection {Neural Network Model}

\NI We consider a recurrent network of $N_{neu}$ model neurons. The total sum of inputs entering neuron $n$ in the discrete time step $t$ is given by

\begin{equation}
u_n^{(t)} = \left(
v_n +
\sum_{m=1}^{N_{neu}} w_{nm} s_m^{(t)} \right) + \left(
q\;x_n^{(t)}
\right) + \left( 
r\;\eta_n^{(t)}
\right).
\label{TotSum}
\end{equation}
Here, the first bracket contains a possible bias $v_n$, as well as a weighted sum of the momentary output signals $s_m^{(t)}$ from all neurons $m$ in the network. The weighting $w_{nm}$ describes the coupling strength from source neuron $m$ to target neuron $n$.  The second bracket accounts for the momentary external input signal $x_n^{(t)}$ entering neuron $n$, scaled by a global input coupling parameter $q$. Finally, the third bracket accounts for a random signal $\eta_n^{(t)}$ entering neuron $n$, scaled by a global {\bf noise strength} parameter $r$. The $\eta_n^{(t)} \sim \mathcal{N}(0, 1)$ are statistically independent random numbers, drawn from a standard Gaussian distribution with mean $0$ and standard deviation $1$. Further on, we will denote the complete {\bf weight matrix} by $\mathbf{W}$, the bias vector by $\mathbf{v}$, the momentary state vector by $\mathbf{s}^{(t)}$, and the external input vector by $\mathbf{x}^{(t)}$. The total sum $u_n^{(t)}$ of inputs, as defined in Eq.~\ref{TotSum}, is used to compute the next output state $s_n^{(t\!+\!1)}$ of neuron $n$. This {\bf update}, which {\bf is performed synchronously} for all neurons,  differs for the two models considered in this paper: 

\NI In the {\bf deterministic model}, neural output signals are continuous in the range $\left[-1,+1\right]$ and are computed directly as the tangens hyperbolicus of the total sum:
\begin{equation}
s_n^{(t\!+\!1)} = \tanh(u_n^{(t)}).
\end{equation}

\NI To initialize the deterministic network, the $N_{neu}$ elements of the zero-time state vector $\mathbf{s}^{t\!=\!0}=\left( s_0^{0},s_1^{0},\ldots,s_{N_{neu}\!-\!1}^{0} \right)$ are drawn independently from a uniform distribution in the range $\left[-1,+1\right]$.\vspace{0.3cm} 

\NI In the {\bf probabilistic model}, neural output signals are discrete with the two possible values $\left\{-1,+1\right\}$. The probability for the value $+1$, also called the 'on'-probability, is computed as a logistic function of the total sum:
\begin{equation}
p_{on,n}^{(t\!+\!1)} = \sigma\left(u_n^{(t)}\right) = \frac{1}{1+\exp(-u_n^{(t)})}.
\end{equation}

\NI To initialize the probabilistic network, the $N_{neu}$ elements of the zero-time state vector $\mathbf{s}^{t\!=\!0}=\left( s_0^{0},s_1^{0},\ldots,s_{N_{neu}\!-\!1}^{0} \right)$ are drawn independently from a Bernoulli distribution, in which the possible outcomes $-1$ and $+1$ occur with equal probability.

\NI We also refer to our probabilistic model as a {\bf Symmetrical Boltzmann Machine (SBM)}, which is called 'symmetric' because the binary outputs are set to $\left\{-1,+1\right\}$, rather than the values $\left\{0,1\right\}$ in conventional Boltzmann machines. Our choice makes the SBMs directly comparable to deterministic tanh-networks with the same weight matrix.

\NI Note that here we do not apply any input to the recurrent neural networks, and thus $q\!=\!0$. Also, we do not use any biases, so that $v_n\!=\!0$ as well. For some types of sparse weight matrices considered in this work, the elements have ternary values $w_{nm} \in \left\{ -w,0,+w \right\}$. In this case, $w$ is called the {\bf weight magnitude} parameter. The same name is also used for a multiplicative parameter $w$ that scales the standard deviation of a weight matrix with originally random normal elements $w_{nm}\sim \mathcal{N}(0,1)$. 

\NI After defining the weight matrix and randomly initializing a network, the time series of global system states $\mathbf{s}^{t\!=\!0}, \mathbf{s}^{t\!=\!1}, \ldots$ is computed numerically for $N_T$ time steps, a parameter also called the observation {\bf time scale}. 

\subsection {Information Theoretic Quantities}

\NI Numerical evaluation of information theoretic quantities requires data with discrete values. The binary output of the SBM is perfectly suited for this purpose, but in the case of the tanh-network we first needed to binarize the continuous outputs $s_n^{(t)}\!\in\!\left[-1,+1\right] \rightarrow x_n^{(t)}\!\in\!\left\{0,1\right\}$, defining $x\!=\!0$ if $s\!<\!0$, and $x\!=\!1$ if $s\!\ge\!0$. Thus, for both types of networks, the output time series can be reduced to a binary matrix $\mathbf{X}\cong\mathbf{s}^{(t)}$, or $\mathbf{Y}\cong\mathbf{s}^{(t\!+\!1)}$ for the time-shifted series. The matrix $\mathbf{X}$ has $N_T$ rows, each with $N_{neu}$ binary values $0$ or $1$. The rows correspond to momentary global states $x$ of the network, and (due to the binarization) the total number of possible global states is $N_x = 2^{N_{neu}}$.

\NI The starting point for all our information theoretic quantities is the {\bf joint probability} $P(x,y)$ that a global system state $x$ is followed by a subsequent state $y$ in the time series. Since the size $N_x$ of the state space increases exponentially with the network size $N_{neu}$, the estimation of $P(x,y)$ becomes quickly unfeasible for large systems: Not only does it take too long before the system has ergodically spread over its entire state space and tried out all possible transitions $x\rightarrow y$, but in large systems it even becomes difficult to hold the huge matrix $P(x,y)$ in the computer memory. To alleviate the memory problem, we hold only the matrix elements between the subset of global system states that have actually been visited by the system within the time scale $N_T$, which can be much smaller if the system is trapped in a dynamical attractor. From the joint probability $P(x,y)$ we directly obtain the two {\bf marginal probabilities} $P(x)$ and $P(y)$, which in our case are practically identical, because each final state becomes an initial state in the next time step. 

\NI The first information theoretical quantity of interest in the {\bf state entropy} $H(X)$ of global system states, defined by
\begin{equation}
H(X) = H(Y) =- \sum_x P(x) \log_2 P(x),
\end{equation}
where all terms with $P(x)\!=\!0$ count as zero.

\NI The next relevant quantity is the {\bf mutual information} $I(X;Y)$ between subsequent system states, defined by
\begin{equation}
I(X;Y) = \sum_x \sum_y P(x,y) \log_2 \left( \frac{P(x,y)}{P(x) P(y)} \right),
\end{equation}
where all terms with $P(x)\!=\!0$, or $P(y)\!=\!0$, or $P(x,y)\!=\!0$ count as zero.

\NI The final important quantity is the conditional entropy $H(Y|X) = H(Y)-I(X;Y)$, which in our case can also be written as $H(Y|X) = H(X)-I(X;Y)$. This conditional entropy $H(Y|X)$ describes the random divergence from a specific initial state $x$ to several possible final states $y$ and is therefore called the {\bf state divergence} $D=H-I$ in the following. A value of $D\!=\!0$ would indicate perfectly deterministic behavior, a value of $D\!=\!H$ perfectly random behavior. 

\section{Results}

\subsection{Conditions and Mechanisms of Recurrence Resonance}

\NI Our first goal is to identify the preconditions of RR, in particular regarding the network weight matrices. For this purpose, we consider a Symmetric Boltzmann Machine (SBM, see Methods for details) with $N_{neu}\!=\!5$ binary neurons. Since such a system has only $N_x\!=\!2^5\!=\!32$ global states $x$, the entropy $H(X)$, the mutual information $I(X;Y)$ between subsequent system states $x$ and $y$, as well as the divergence $D\!=\!H\!-\!I$ can be accurately estimated from the network's output time series on a manageable time scale of $N_T=10^4$ time steps (See Methods for details).

\NI As a basis for the network's weight matrix $\mathbf{W}$, the 25 elements are first drawn independently from a standard normal distribution $\sim \mathcal{N}(0,1)$ (Fig.\ref{fig_GW}(a)). This 'frozen random matrix' is then multiplied (scaled) with a weight magnitude parameter $w$, thus tuning the standard deviation of the matrix, while keeping its fundamental structure invariant. We can then explore how the three information theoretic quantities $H$, $I$ and $D$ depend on the weight magnitude $w$ and on the strength $r$ of added white noise (See Methods for details).

\NI Before that, it is useful to imagine the dynamical structure of the SBM as a state transition graph, in which the 32 nodes represent the possible global states $x$, and the weighted directional edges represent the possible transitions $x\rightarrow y$ between pairs of states. The probability of each transition, which is given by the conditional probability $P(y|x)$, may be indicated by the thickness of the edges. This state transition graph, which is determined only by the weight matrix $\mathbf{W}$, describes the {\em possible} dynamical behavior of the network completely, irrespective of which path the system is actually taking through the graph. 

\NI Below, we will visualize the aggregated activity in the network by the joint probability $P(y,x)=P(y|x)P(x)$. While this quantity also respects the fundamental transition possibilities (described by $P(y|x)$), it additionally accounts for which states and transitions the system has {\em actually} used during the observation time (described by $P(x)$).

\subsubsection{System behavior without noise}

\NI We first consider the system without applying external noise ($r\!=\!0$). For a weight magnitude of $w=0$, the five neurons are completely isolated from each other and also have no self-connections (autapses). Consequently, each of the probabilistic SBM neurons produces a non-biased, temporally uncorrelated, binary random walk, where the two possible values $-1$ and $+1$ occur with equal probabilities, and independently from each other. This means that any momentary global system state $x$ can transition to any other global successor state $y$ with equal probability. We thus have a 'structureless', fully connected state transition graph, with equally 'thick' edges. 

\NI Since the time scale of $N_T\!=\!10^4$ is long enough for the system to explore its entire space of $N_x\!=\!32$ global states, we expect that for $w\!=\!0$ the entropy reaches its maximum possible value $H\!=\!N_{neu}\!=\!5$. As the system is purely random with a structureless state transition graph, the mutual information is expected to be $I\!=\!0$. Consequently, the divergence $D=H-I$ is also maximal at $D\!=\!N_{neu}\!=\!5$. This is indeed found in the numerical simulation.

\NI As we tune the weight magnitude $w$ from zero to increasingly positive values, the state transition graph, being determined by the weight matrix, is gradually developing a structure, that is, some of the transitions become more probable than others. Consequently, certain dominating paths are forming within the graph along the 'thick' edges, leading eventually to the emergence of dynamical attractors, such as fixed points (1-cycles), higher period n-cycles, or transient states. These attractors are quite unstable at low weight magnitudes $w$, so that the system can still transition between them. Nevertheless, within the limited time horizon of $N_T$ time steps, it now becomes impossible to visit all 32 global states (nodes) with equal probability. Without additional noise ($r\!=\!0$), this leads to a gradual decrease of  the entropy $H(r\!=\!0)$ (Fig.\ref{fig_GW}(b)). The growth of structure in the graph with increasing $w$ makes the system dynamics more deterministic and is thus also connected with a decrease of the random divergence $D(r\!=\!5)$ (not shown in the figure). 

\NI However, since the entropy $H$ and the divergence $D$ decrease at different rates with the weight magnitude $w$, the mutual information $I=H-D$ shows a more complicated behavior (Fig.\ref{fig_GW}(c)). As $w$ is increased from 0.2 to 2, the mutual information without noise $I(r\!=\!0)$ is first increasing, reflecting the higher degree of predictability of the next state. But for $w\!=\!5$, we find that $I(r\!=\!0)$ is decreasing again, reaching a value of zero for $w\!=\!10$ (Fig.\ref{fig_GW}(c), red and magenta curve).

\NI This extreme situation of $H\!=\!D\!=\!I\!=\!0$, which is most often found at large weight magnitudes but without noise, means that the system is trapped in a single global state, in other words, a quasi-stable fixed point $x^{\star}$. Correspondingly, the joint probability $P(x,y)$ of subsequent system states has only a single non-zero entry at $x\!=\!y\!=\!x^{\star}$ (Fig.\ref{fig_GW}(d,matrix plot)), and also the marginal state probability $P(x)$ has only a single entry at the fixed point state $x\!=\!x^{\star}$ (Fig.\ref{fig_GW}(d,histogram on top)). 

\subsubsection{System behavior with noise}

\begin{figure}[ht!]
\centering
\includegraphics[width=1\linewidth]{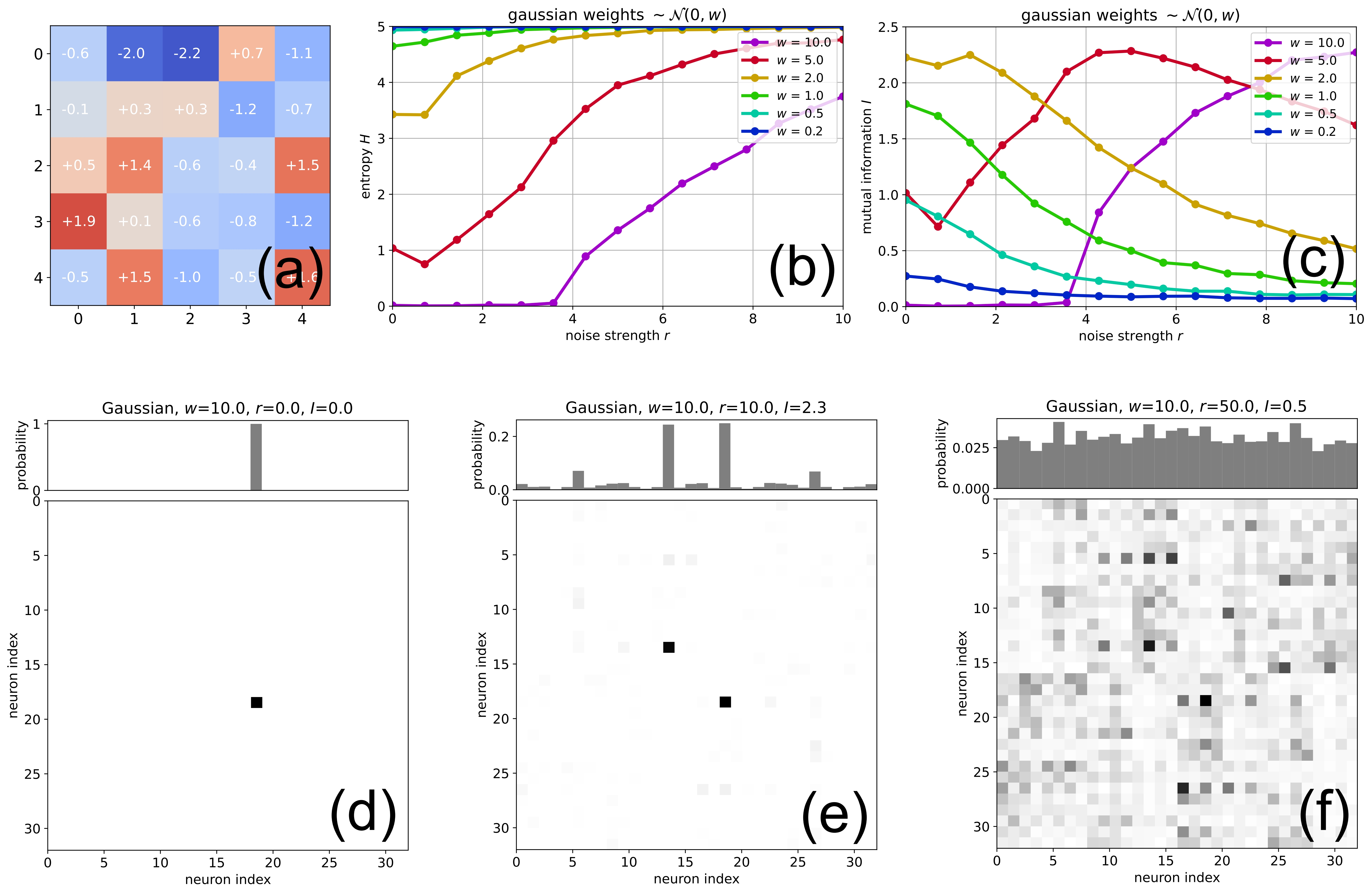}
\caption{{\bf Effect of weight magnitude on Recurrence Resonance in Boltzmann Machines:} 
{\bf(a)} The elements of a 5$\times$5 weight matrix are drawn randomly from a standard normal distribution $\sim \mathcal{N}(0,1)$. The matrix is then scaled (multiplied) by a tunable weight magnitude parameter $w$, thus changing the standard deviation of the elements while keeping the fundamental structure of the matrix fixed. 
{\bf(b)} The entropy $H$ of global RNN states as a function of the strength $r$ of added noise, for six different $w$. For very small weights $w\!\leq\!1$ (dark blue, light blue and green curves) the system has near its maximum possible entropy $H\!=\!5$, as the neurons fire almost independently. The entropy $H$ at $r\!=\!0$ is however drastically reduced as the weights get stronger ($w\!>\!1$, orange, red and magenta curves), because the system is then trapped within a single attractor. In this trapped state, application of a sufficient level of noise $r\!>\!0$ allows the system to escape and visit other attractors as well, leading to an increase of the entropy.
{\bf(c)} The mutual information $I$ between subsequent global RNN states as a function of the strength $r$ of added noise, for six different $w$. In the regime of low weights $w$, adding noise leads to a decrease of the mutual information $I$. However, for sufficiently large $w$ (red and magenta curves), the mutual information $I$ increases with added noise $r$ and shows a maximum at some optimal noise level (at around $r_{opt}\approx 5$ for the red curve).
{\bf(d)} The joint probability $P(\mathbf{s}^{(t)},\mathbf{s}^{(t\!+\!1)})$ of subsequent global system states, for the RNN with width magnitude parameter $r=10$, without noise. The system is trapped in one of the 32 possible states, which is also visible in the marginal distribution $p(\mathbf{s}^{(t)})$ of system states (top of the matrix plot). This resting in the fixpoint corresponds to a mutual information of $I\!=\!0$.
{\bf(e)} Adding a noise of strength $r\!=\!10$ allows the system to visit another quasi-stable fixpoint (as well as some other transient states) for a fraction of time steps, increasing the mutual information to $I\!=\!2.3$.
{\bf(f)} Adding a too large level $r\!=\!50$ of noise opens up almost all possible states and transitions for the system. However, the system can no longer stay within any particular attractor for a longer period of time. This loss of order corresponds to a drop of the mutual information to $I\!=\!0.5$. 
} 
\label{fig_GW}
\end{figure}

\NI We now go back to the case of relatively weak weight magnitudes $w\leq1$ and gradually increase the strength $r$ of added noise. Here, the entropy $H$ is already quite large without noise, and adding noise increases it even further (Fig.\ref{fig_GW}(b, green curves on the very top)). The noise thus helps the system to visit all possible states with equal probability, which can be seen as a beneficial effect. However, the noise increases also the divergence $D$ at a fast rate (not shown in the figures), so that the mutual information $I=H-D$ is only decreasing as more noise is added (Fig.\ref{fig_GW}(c, green, light blue and dark blue curves)). We therefore do not observe RR in the regime of weak weight magnitudes.

\NI In contrast, a different behavior is found for stronger weight magnitudes $w\geq5$. Here, the entropy $H$ is also increased by adding more noise (Fig.\ref{fig_GW}(b, red and magenta curves)), but the mutual information $I$ is now initially increasing with noise - with a small exception at small noise levels (Fig.\ref{fig_GW}(c, red and magenta curves)). In the case of $w\!=\!5$ (red curve), it reaches a maximum at a noise level of around $r\!=\!5$ and then decreases again. For $w\!=\!10$ (magenta curve), the maximum of $I$ is around $r\!=\!10$. Thus, RR is only observed in the regime of sufficiently strong weight magnitudes.

\NI Generally, whenever the mutual information as a function of noise shows a clear maximum in a given network, the joint and marginal probability distributions are characteristically different at the points without noise (Fig.\ref{fig_GW}(d), $r\!=\!0$), close to the RR maximum (Fig.\ref{fig_GW}(e), $r\!=\!10$), and far beyond the RR maximum (Fig.\ref{fig_GW}(e), $r\!=\!50$): Without noise, the system visits relatively few states, spending its time in only one or a few attractors. At the RR peak, the number of visited states is larger, and those states mainly belong to quasi-stable attractors (The matrix plot then shows a distinct set of dominating entries). Beyond the RR peak, the system visits even more states, but now only transiently, without staying in any particular attractor for a longer period of time (The matrix plot then appears more uniform and unstructured than at the RR maximum).

\subsection{RR in Selected Networks with Multiple Attractors}

\NI If RR is a process where noise helps neural networks to reach more attractors in a given time horizon, the phenomenon should be particularly pronounced in systems with multiple (as well as sufficiently stable) attractors. We therefore select the following three specific types of weight matrices, while keeping the network size of the SBM at $N_{neu}\!=\!5$ and the time scale at $N_T\!=\!10^4$: an autapses-only network, a Hopfield network, and a NRooks network.

\subsubsection{Autapses-only network}

\begin{figure}[ht!]
\centering
\includegraphics[width=1\linewidth]{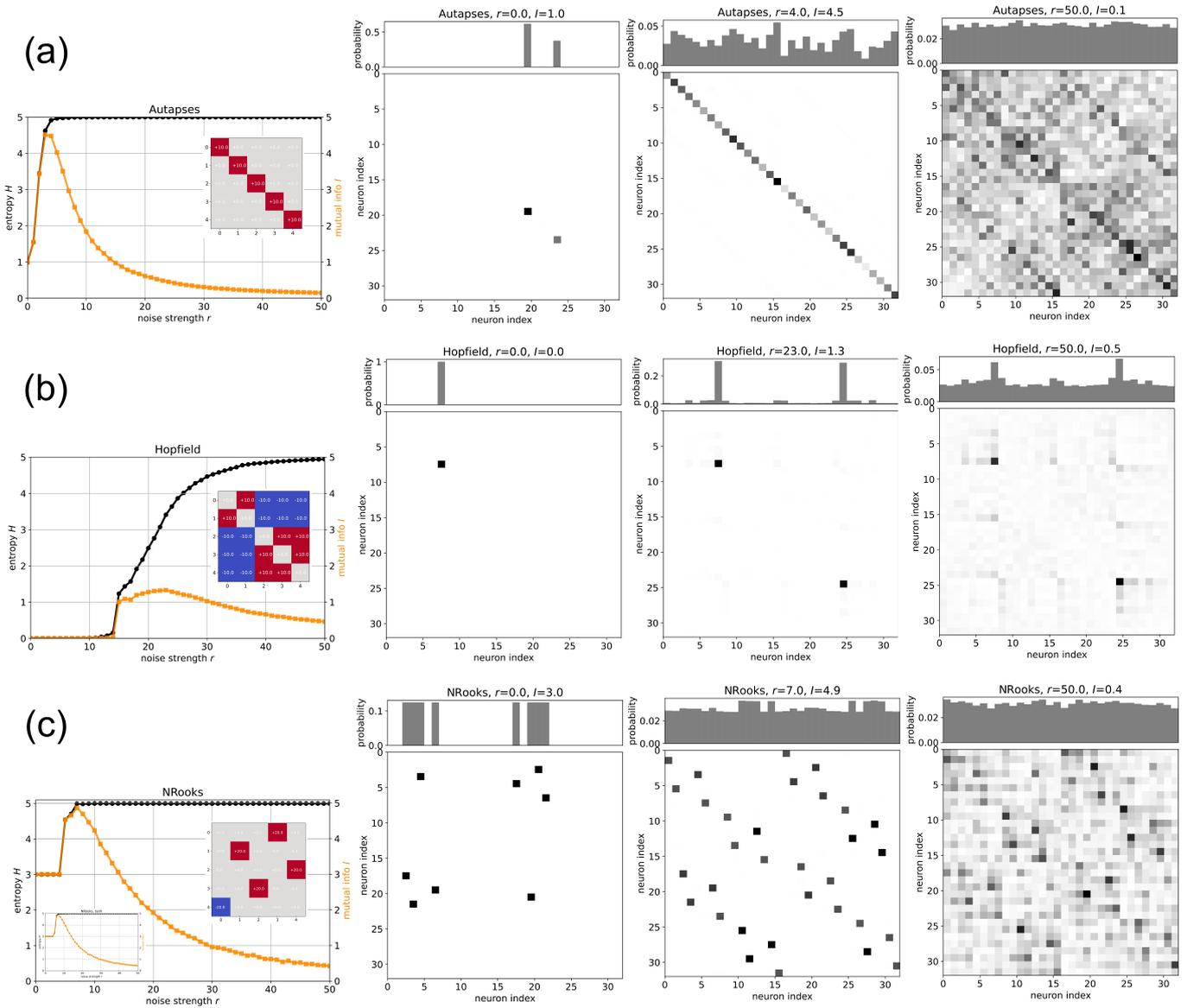}
\caption{{\bf Recurrence Resonance in various multi-attractor RNNs:} We consider three types (rows a,b,c) of Boltzmann Machines with 5 neurons. The first column in each row is a plot of the entropy $H$ (black) and of the mutual information $I$ (orange) as a function of the noise strength $r$, with the weight matrix as an inset. Columns two to four show the joint probability matrices $P(\mathbf{s}^{(t)},\mathbf{s}^{(t\!+\!1)})$ of subsequent global system states, with the marginal state distributions $p(\mathbf{s}^{(t)})$ on top, for the zero noise case $r\!=\!0$, optimal noise $r\!=\!r_{opt}$, and excessive noise $r\!=\!50$. 
{\bf(a)} Unconnected neurons with strong positive {\bf autapses} (self-connections).
{\bf(b)} {\bf Hopfield} network, designed to store two distinct patterns (fixed points).
{\bf(c)} {\bf NRooks} networks, in which each column and row contains only one non-zero weight matrix element of large magnitude. The left inset corresponds to a network with the same weight matrix, but with deterministic tanh-neurons.
} 
\label{fig_MA}
\end{figure}

\NI We first test a {\bf 'autapses-only' network}, in which all non-diagonal elements of  the weight matrix (corresponding to inter-neuron connections) are zero, whereas the diagonal elements (corresponding to neuron self-connections, or 'autapses') have the same positive value $w=+10$ (Fig.\ref{fig_MA}(row (a), inset of left plot)). In this case, since the SBM neurons are isolated from each other, they produce mutually independent, binary random walks. The probabilities of $-1$ and $+1$ are still equal, because we do not use biases ($v_n\!=\!0$, See Methods). However, due to the excitatory autapses, the random walks are now temporally correlated ('persistent'), which means that an output of $+1$ is more likely followed by another $+1$, and analogously for $-1$. Each neuron thus tends to produce longer chains of outputs with identical sign, only switching to the opposite sign after a certain correlation time. For the total system, this means that any momentary global state is conserved, with high probability, for a finite number of time steps. Hence, each of the 32 global system states is a quasi-stable fixed point here, and the degree of stability can be increased by the self-connection strength $w$. For this reason, this type of network could be useful as short-term memories in practical applications.

\NI In our simulation, the autapse-only network without external noise is spending all of the $10^4$ time steps in only two of its 32 fixed point attractors, consequently leading to a mutual information of about $I\!=\!1$ (Fig.\ref{fig_MA}(row (a), column 'no noise')). Adding an optimal amount of noise ($r\!=\!4$) drives the network to visit all available attractors, yet not with the same frequency. Nevertheless, the peak mutual information is with $I\!=\!4.5$ close to the upper limit of 5 (Fig.\ref{fig_MA}(row (a), column 'optim. noise')). Applying an excessive noise of $r\!=\!50$ lets the system undergo almost all $32^2$ possible state-to-state transitions, but these frequent unpredictable jumps between attractors lead to a mutual information of only $I\!=\!0.1$ (Fig.\ref{fig_MA}(row (a), column 'strong noise'))

\NI A remarkable feature of the autapse-only network's RR-curve (Fig.\ref{fig_MA}(row (a)) is the part between zero and optimal noise. In this regime, the mutual information $I$ follows extremely closely  the rising curve of the entropy $H$, meaning that the divergence $D$ is extremely small. Hence, the noise is on the one hand able to occasionally transfer the system from one (fixed point) attractor to a different one, but on the other hand allows the system to stay for a sufficiently long time interval within each attractor, so that the next state remains predictable to a high degree which constitutes a precondition for a high mutual information. We will see below that other systems also show this initial noise regime where entropy and mutual information rise together, while the divergence remains close to zero. At some level of noise, of course, the divergence must increase as well.

\subsubsection{Hopfield network}

\NI Another type of recurrent neural network that is famous for its ability to have multiple (designable) fixed point attractors is the {\bf Hopfield network} \cite{hopfield1982neural}. Weight matrices of Hopfield networks are symmetric ($w_{mn}\!=\!w_{nm}$) and have no self-connections ($w_{mm}\!=\!0$). Its neurons are traditionally updated one by one in an asynchronous manner, but we continue to use a synchronous update in our SBM model. 

\NI We have designed the weight matrix to 'store' the two patterns $\;[+1+1-1-1-1]\equiv24\;$ and $\;[-1-1+1+1+1]\equiv7\;$. The magnitude $w$ of the weight matrix elements was made large enough to ensure a good stability of the two fixed points corresponding to the stored patterns (Fig.\ref{fig_MA}(row (b), inset of left plot)). 

\NI In a broad initial regime of noise strengths $r\!=\!0\ldots10$, the RR-curve of the Hopfield network (Fig.\ref{fig_MA}(row (b), left plot)) shows an entropy $H$ and mutual information $I$ of zero, which is characteristic for a system that is trapped in a single fixed point. The matrix plot of the joint probability reveals that this fixed point is the global state $\;[+1+1-1-1-1]\equiv24\;$, the first of the two stored patterns (Fig.\ref{fig_MA}(row (b), column 'no noise')). The mutual information $I$ starts to rise sharply at around $r\!=\!15$ and reaches a peak at $r\!=\!23$, while the entropy $H$ continues to increase toward the upper limit. At the optimal noise level, the system is now visiting both fixed points (state $7$ as well as state $24$) with similar frequency, resulting in a peak mutual information of about $I\!=\!1.3$ (Fig.\ref{fig_MA}(row (b), column 'optim. noise')). Since the two fixed points are very stable (due to the large magnitude of matrix elements), the system is spending a comparatively large fraction of time in these two states, even at a large noise level of $r\!=\!50$ (Fig.\ref{fig_MA}(row (b), column 'strong noise')).

\subsubsection{NRooks network}

\NI Finally, we test the RR phenomenon in so-called {\bf 'NRooks' networks}, which under ideal conditions (large magnitude $w$ of non-zero weights and long observation time scale $N_T$) are known to approach the upper theoretical limit of mutual information and entropy, corresponding to $I\!=\!H\!=\!N_{neu}$ \cite{metzner2024quantifying}, and zero Divergence $D\!=\!0$. The NRooks weight matrix has only one non-zero matrix element in each row and column (hence the name), and these $N_{neu}$ non-zero matrix elements have the same magnitude $w$, but arbitrary signs (Fig.\ref{fig_MA}(row (c), upper inset of left plot)). It has been shown that all global states of an NRooks system are parts of n-cyles of various sizes, that is, there are no transient states that would merely lead into these attractors \cite{metzner2024quantifying}.

\NI Our specific NRooks system turns out to have four different 8-cycles as attractors, and without noise it is trapped in one of them (Fig.\ref{fig_MA}(row (c), column 'no noise')). Since running for thousands of time steps within this attractor involves 8 distinct states (creating entropy) in a perfectly predictable order (without divergence $D$), the entropy and mutual information have already a relatively large value of $H\!=\!I\!=\!3$, even without external noise (Fig.\ref{fig_MA}(row (c), left plot)). Since attractors are quite stable at $w\!=\!20$, we observe a plateau with $H\!=\!I\!=\!3$ in the RR-curve, holding up to a noise level of $r\!=\!4$, where $H$ and $I$ start to increase rapidly. At the peak of the mutual information, occurring for a noise level of $r\!=\!7$, it reaches the value of $I\!=\!4.9$, which is close to the theoretical maximum of $5$. Indeed, at that point the system is visiting all four 8-cycle attractors with about the same frequency (Fig.\ref{fig_MA}(row (c), column 'optim. noise')). It stays for a very long time in each of them, behaving almost perfectly deterministic. Only occasionally, the noise of optimal strength 'kicks' the system randomly to one of the other three attractors. As usual, for a very strong amount of noise, the system looses its predictability, and the mutual information $I$ drops accordingly, while the entropy $H$ remains at the upper limit.

\NI In the above numerical experiments with multi-attractor SBMs, we have used relatively large weight magnitudes $w$. In the given context, this served the purpose to make the attractors of the autonomous networks more stable. More generally, large weight magnitudes drive the SBM neurons into the saturation regime of the logistic activation function, so that the on-probabilities $p_{on,n}^{(t)}$ become either $\approx\!0$ or $\approx\!1$ for all neurons $n$ and all time steps $t$. Hence, the probabilistic SBM then behaves quasi deterministic. For this reason, we could use an SBM to implement a Hopfield network, which is usually based on deterministic binary threshold neurons.

\NI In order to further demonstrate the saturation regime of the SBM, we have used the same weight matrix that was used in the NRooks example also in a network with deterministic tanh-Neurons (See Methods for details). The resulting RR-curve is indeed extremely similar to that of the probabilistic SBM (Fig.\ref{fig_MA}(row (c), lower inset of left plot)).

\subsection{Time-scale dependence of RR}

As already mentioned above, the observation time scale $N_T$ is another critical factor that determines whether a RR peak will be observable in a given system.

\begin{figure}[ht!]
\centering
\includegraphics[width=0.6\linewidth]{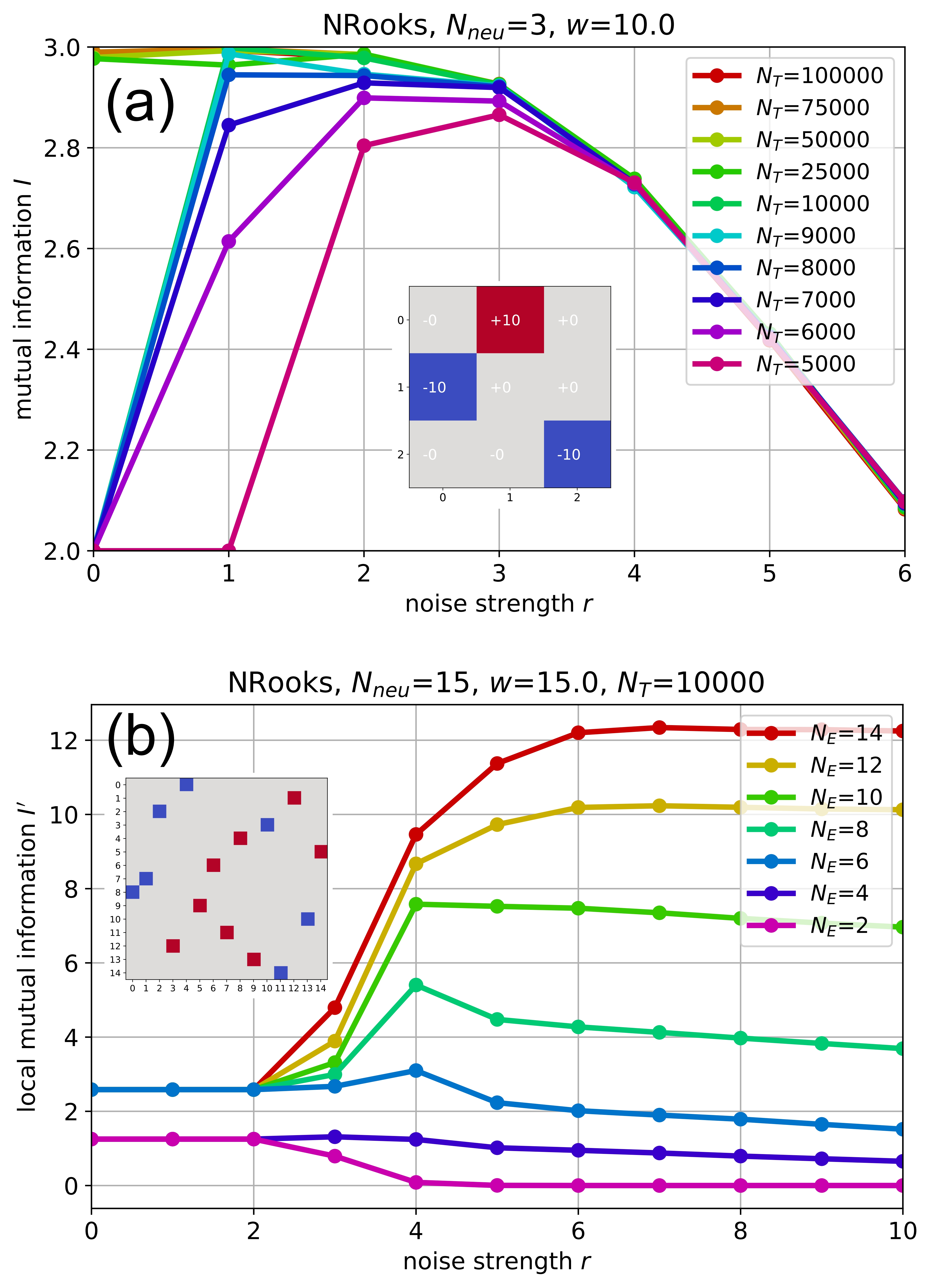}
\caption{
{\bf (a): Time-scale dependence of information quantities.}
Mutual information $I$ in a 3-neuron NRooks system, with weight magnitude $w\!=\!10$ of the non-zero matrix elements, as a function of the noise strength $r$, evaluated for different numbers of simulation time steps $N_T$ between 5000 and 100000. All curves show the same decay of mutual information $I$ for noise levels larger than about $r\!=\!4$, but they differ drastically in the regime of smaller noise levels: For a proper time scale $N_T\!=\!5000$, the curve $I\!=\!I(r)$ shows clearly the RR phenomenon, with a peak mutual information at $r\!=\!3$. As the observation time scale $N_T$ is increased, the mutual information $I$ in the whole regime of smaller noise levels is rising, as the system gets more opportunities to switch attractors. This leads to a shift of the RR peak to lower noised levels $r_{opt}$. For time scale above $N_T\!\approx\!10000$, the mutual information $I(r\!=\!0)$ without noise is rather abruptly jumping to close to the maximum possible value, meaning that the system does not require external help any more to realize optimal information flux. At this point the RR phenomenon disappears. {\bf (b): 'Local' mutual information $I^{\prime}$} in a 15-neuron NRooks system with weight magnitude $w\!=\!15$ and time scale $N_T\!=\!10^4$, as a function of the noise strength $r$, evaluated only in a subgroup of $N_E\in \left\{2,4,6,8,10,12,14\right\}$ neurons. Since the time scale of $N_T\!=\!10^4$ is insufficient to reach 'ergodic' behaviour in a space of $2^{15}$ states, no RR maximum is visible for large subgroup sizes $N_E\!=\!12$ and $N_E\!=\!10$, but instead a rise and saturation of $I^{\prime}$ with increasing noise (plateau remains until at least $r\!=\!50$, data not shown). For too small subgroup sizes $N_E\!=\!2$ and $N_E\!=\!4$, only a monotonous drop of mutual information is observed as a function of the noise level $r$. However, for intermediate subgroup size $N_E\!=\!6$ and $N_E\!=\!8$, a clear RR peak is emerging. It is therefore possible to detect RR even in large systems, at least locally.}
\label{fig_TS}
\end{figure}

\NI For a demonstration, we now use a NRooks system of only 3 neurons (Fig.\ref{fig_TS}(a)). Because of its extremely small state space ($N_x\!=\!8$), it is possible to make the attractors stable by a relatively large weight magnitude of $w\!=\!10$, but nevertheless to approach the ergodic limit of very large time scales $N_T$, where the system is able to visit all its attractors autonomously, without the injection of external noise.

\NI We find that for too strong levels of noise (here $r\!\geq\!4$), the mutual information $I$ is just declining, irrespective from the time scale $N_T$. In this regime, the system is operating already at maximum entropy ($H\!\approx\!3$), but the noise is causing an increasing loss of predictability.

\NI For a time scale of $N_T\!=\!5000$, which is appropriate for the given system, a clear RR peak is observed in the mutual information curve $I\!=\!I(r)$ at around $r\!=\!3$. As the time scale is now prolonged, up to around $N_T\!=\!9000$, the mutual information $I\!=\!I(r)$ is generally rising for moderate noise levels in the range $r\in\left[0,3\right]$, because the systems gets more opportunities to escape and switch from one attractor state to another. As a consequence, the RR peak is moving to smaller noise levels $r_{opt}$.

\NI For the moderate time scales $N_T\!<\!9000$ considered so far, the system, without noise, is trapped in a 4-cycle, and therefore the mutual information is $I(r\!=\!0)=2$. However for larger time scales $N_T\!>\!10000$, we find a value $I(r\!=\!0)\approx3$ close to the theoretical maximum of $3$, which means that now even the zero-noise system can visit all its attractor states and run through each of them in an almost perfectly predictable way. Thus, the phenomenon of RR is not observable on extremely long time scales, where systems already operate close to their ergodic regime.

\subsection{'Local' mutual information in sub-networks}

In actual applications of RNNs, such as reservoir computing, networks are typically so large ($N_{neu}\!>\!100$) and consequently the state spaces so huge ($N_x\!>\!2^{100}$) that the ergodic regime cannot be reached on any practical time scale $N_T$. Moreover, in such practically non-ergodic systems, it is also impossible to accurately evaluate the full mutual information of subsequent system states, because the joint probability matrices are too large ($N_x \times N_x \!>\!2^{100} \times 2^{100}$) and because the empirical distributions have not enough time to converge toward a stable result. The question then arises how to compute a useful approximation of $I\!=\!I(r)$ in large systems, even when they are observed on 'too short' (but practically relevant) time scales. 

\NI Although a detailed investigation of this question is beyond the scope of the present paper, we provide a first insight using a 15-neuron NRooks system, observed on the non-ergodic time scale of $N_T\!=\!10000$ (Fig.\ref{fig_TS}(b)). To alleviate the matrix size problem, we only consider global states that have actually been visited during the observation time (See Methods for details). Moreover, as a proxy for the full $I(r)$, we compute the 'local' mutual information $I^{\prime}(r)$ within smaller sub-networks, i.e. subgroups of only $N_E\!\leq\!N_{neu}$ neurons.

\NI For large sub-networks ($N_E\!>\!10$), instead of a RR peak, we find an increase of the local mutual information $I^{\prime}(r)$ with noise, and finally a saturation. This plateau is also observed for larger noise levels up to $r\!=\!50$ (data not shown).

\NI In contrast, for very small sub-networks ($N_E\leq4$), we find already a smaller starting value of $I^{\prime}(r\!=\!0)$ at zero noise, and eventually a decline of $I^{\prime}(r)$ with increasing noise level $r$.

\NI However, for a certain intermediate range of sub-network sizes between $N_E\!=\!6$ and $N_E\!=\!8$, the curve $I^{\prime}(r)$ shows a clear maximum that decays very slowly after the peak. Thus, in large networks, when observed on short quasi non-ergodic time scales, a phenomenon similar to RR can occur within smaller sub-networks, whereas the mutual information of the total system then shows a saturation-type dependence on the noise level. 

\subsection{Controlling probabilistic, binary-valued RNNs by noise pulses}

At the peak of the RR curve, the continuous white noise input of optimal strength $r_{opt}$ is leaving a RNN in its present attractor for long times, but occasionally causes a random transit to one of the other available attractors. It is this combination of high predictability and high entropy that leads to the optimal value of the mutual information.

\begin{figure}[ht!]
\centering
\includegraphics[width=1\linewidth]{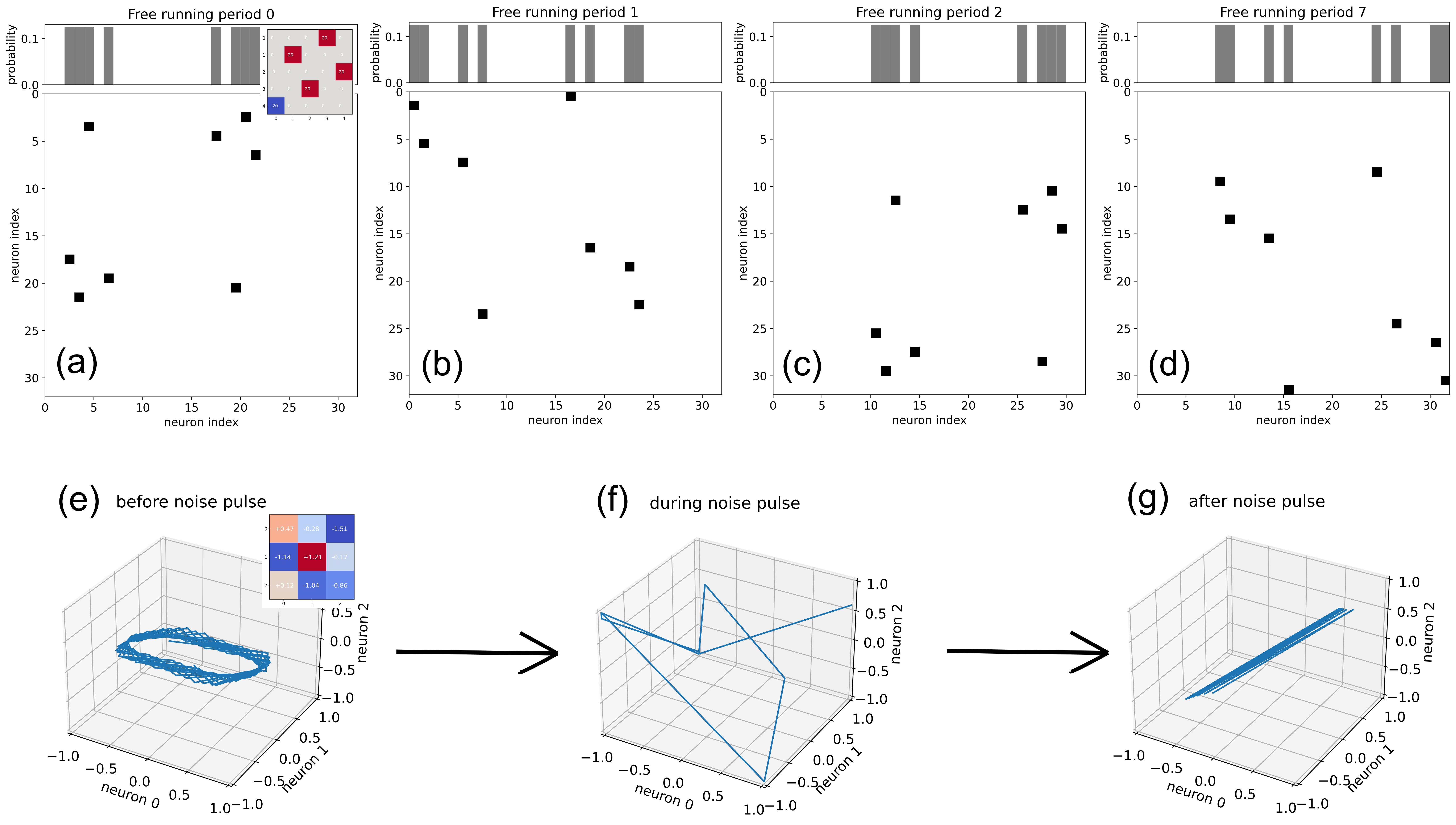}
\caption{ {\bf Controlling RNNs by noise pulses.} {\bf (a-d): 5-neuron NRooks system with a state space consisting of four 8-cycles}. The system (weight matrix see inset) is originally trapped in one of these attractors (a). After applying a noise pulse with a duration of 10 steps and a strength of $r\!=\!50$, the system has been randomly transferred to another 8-cycle (b). A further noise pulse moves the system into the third available 8-cycle (c). The next four noise pulses produce the same attractors that have been already visited, but the subsequent noise pulse transfers the system into the final of the four distinct attractors (d). Without adding noise, the system remains stable in its current attractor for arbitrarily long times, as determined by the weight magnitude of the non-zero weights (here $w\!=\!20$). {\bf (a-d): 3-neuron system with random Gaussian weight matrix and deterministic tanh-neurons}. Plotted is the trajectory of the system (weight matrix see inset) in the continuous state space cube $\left]-1,+1  \right[^3$. Originally, the system is running through a strange attractor, where the trajectory is approximately confined within a torus (e). After 100 steps of free time evolution, a noise pulse is applied with a duration of 5 steps and a strength of $r\!=\!5$ (f). After the pulse, the system is in a new attractor resembling a 2-cycle, but without precise re-visitation of the endpoints. These examples show that short noise pulses may be a way to access, on demand, various new attractors of a system, but without compromising the perfect order within each attractor.
}
\label{fig_NP}
\end{figure}

\NI A natural extension (and putative application) of this concept are short noise pulses - applied only at times when a change of attractor state is required - instead of a continuous feed-in of noise. To test this concept, we have again used the 5-neuron NRooks system of Fig.\ref{fig_MA}(c), with its four different 8-cycles as attractors. The system is initially in one of its 8-cycle attractors (Fig.\ref{fig_NP}(a)), and remains in this attractor for an arbitrarily long period that only depends on the weight magnitude $w$. By applying short (10 time steps) yet strong ($r\!=\!50$) Gaussian white noise pulses, we could indeed transfer the system randomly to one of the other attractors. It also happens that the system ends up in the same attractor (yet at a different 'phase' of the periodic cycle), but eventually we could reach all four 8-cycles by this way (Fig.\ref{fig_NP}(b-d)).

\subsection{Controlling deterministic, continuous-valued RNNs by noise pulses}

So far, we have only briefly explored the effect of noise on networks of deterministic tanh-neurons (inset of Fig.\ref{fig_MA}(c)). As a further glimpse into this alternative field of research, we apply a short (5 time steps) and weak ($r\!=\!5$) noise pulse to a very small (3 neurons) tanh-network, in which the 9 matrix elements have been drawn randomly from a standard normal distribution $\sim \mathcal{N}(0,1)$. 

\NI Before the noise pulse, the system is allowed to run freely for 100 time steps. The resulting system states at each time step are here continuous points within the three-dimensional cube $\left]-1,+1  \right[^3$ and can thus be visualized directly as a 3-dimensional trajectory (Fig.\ref{fig_NP}(e-f)). We find the system initially within a 'strange', loop-like attractor (e). During the short noise pulse, the trajectory is erratic and reaches the borders of the state space cube (f). After the pulse, the system has settled in a new strange attractor, which resembles a 2-cycle, but only with an approximate return to the end points in each oscillation period. Thus, it is also possible to achieve a switch of attractor states in deterministic RNNs with continous output values by the injection of noise pulses.

\newpage
\section{Discussion}

\NI In this work, we have re-considered the phenomenon of Recurrence Resonance (RR), i.e. the peak-like dependence of an RNN's internal information flux on the level $r$ of white noise added to each of the neurons \cite{krauss2019recurrence}. The information flux is measured by the mutual information $I\!=\!H\!-\!D$ between subsequent system states, a quantity that grows as more states become available (larger entropy $H$), and/or when each successor state can be better predicted from its predecessor (smaller divergence $D$).

\NI We have shown that a resonance-like peak of $I(r)$ can only be observed in networks that fundamentally have a whole set of relatively stable dynamical attractors available, but which - without external intervention - would remain trapped in one of them during the entire observation time scale $N_T$. In this situation, adding a small level $r\leq r_{opt}$ of noise helps such networks to occasionally jump out of the present attractor and switch into another one, without significantly reducing the predictability of the state sequence within each of these quasi-stable attractors (strong increase of $H$, but weak increase of $D$). If the noise level $r$ is however increased beyond the optimal point $r_{opt}$, predictability is lost and consequently the information flux $I(r)$ is declining again, while the entropy $H$ is still increasing toward its upper limit. By contrast, networks that already have a high internal information flux from the beginning will not show a RR peak, but only a decline of $I$ as a function of $r$.

\NI We have demonstrated the RR phenomenon using Symmetric Boltzmann Machines (SBMs) with different types of weight matrices, including random Gaussian matrices, diagonal matrices (autapse-only networks), Hopfield networks trained on specific patterns, and in NRooks systems that are known to reach the upper limit of information flux. In each case, we demonstrated that the network without noise is trapped in a single or few attractors, based on the joint probability of subsequent system states. An optimal level of noise makes more (or even all) attractors available without too many unpredictable transitions. However, excessive levels of noise cause more or less random jumps between all possible pairs of states. In systems with a very high stability of attractors (induced by a large weight magnitude $w$), we have found that $I(r)$ remains constant at the initial value $I(r\!=\!0)$ for a certain range of noise levels, before it abruptly rises in the way of a phase transition.

\NI We have also demonstrated that RR can only be observed in appropriate time scales $N_T$, relative to the total number of possible system states $N_x=2^{N_{neu}}$. For arbitrarily long observation times (which are of theoretical interest but not so much of practical relevance), a neural network operating in the quasi-deterministic regime, but with at least a small probabilistic component (like SBMs with a large but finite weight magnitude $w$) can eventually visit all its attractors in an ergodic manner, but still stay in each of them for extended time intervals. Then $I(r\!=\!0)$ is already close to the optimum value and application of noise only degrades the information flux.

\NI An interesting problem arises therefore in networks with many neurons and thus an exponentially large state space, such as reservoir computers, or brains. Such systems will necessarily spend all their lives within a negligible fraction of the fundamentally available state space, possibly consisting of only a tiny subset of attractors. One way to cope with this 'practical non-ergodicity' would be a repeated active switching between attractor subsets (perhaps using noise pulses), until a useful one is found, and then to stay there. Alternatively, the networks may be designed (or optimized) such that the useful attractors have a very large basin of attraction. A similar problem has been discussed in the context of protein folding with the 'Levinthal paradox', where naturally existing proteins fold into the desired conformation much faster than expected by a random thermal search in conformation space \cite{zwanzig1992levinthal,karplus1997levinthal,honig1999protein}, probably due to funnel-type energy landscapes \cite{bryngelson1995funnels,martinez2014introducing,wolynes2015evolution,roder2019energy}.

\NI In our context of the RR phenomenon, practical non-ergodicity makes it impossible to compute the stationary information flux in a large network, because the system never reaches a stationary state (marked by constant probability distributions) within any practical time scale $N_T$. When the information flux in a network is evaluated naively, using the 'transient' (not yet converged) joint probability distributions, we have found that $I(r)$ shows a saturating behavior, rather than a maximum. Nevertheless, the 'local' mutual information, evaluated for a suitably sized sub-network, can then still show a RR peak.

\NI Finally, we have explored the repeated application of short noise pulses, rather than feeding continuous noise into the neurons. We could demonstrate that each pulse offers the network a chance to switch to a new random attractor (such as an n-cycle), while the intermediate free running phases allow the system to deterministically and thus predictably follow the fixed order of states within each of the attractors. Based on this noise-induced random switching mechanism, an evolutionary optimization algorithm could be implemented in a recurrent neural network, in which various attractors are tried until one turns out useful for a given task. We speculate that this principle might be used in central pattern generators of biological brains \cite{hooper2000central}, for example in order to find temporal activation patterns for certain motor tasks \cite{marder2001central}. 

\NI In free running SBMs, a neuron's probability of being 'on' in the next time step is computed by a logistic activation function $\sigma(u)$, where $u$ is the weighted sum of inputs from other neurons. Adding white, normally distributed noise to $u$ corresponds to a convolution of the logistic function with a Gaussian kernel, resulting in a 'broadening' of the activation function. Increasing the noise level thus has an effect similar to turning up the 'temperature' parameter $T$ in a re-scaled activation function $\sigma(u/T)$. This opens up a new interpretation of the RR phenomenon in terms of statistical physics, in particular if an energy $E(\mathbf{s}) = -\sum_{mn} w_{mn} s_m s_n$ can be assigned to each global system state $\mathbf{s}$.

In this case, the system - after sufficiently long time - would come to thermal equilibrium, and the probability of finding it at any state $\mathbf{s}$ would be proportional to the Boltzmann distribution $\propto\!e^{-E(\mathbf{s})/T}$. At low temperature, the system would therefore spend most of its time in the deepest valleys of the energy landscape, which may correspond to fixed point attractors. Increasing the temperature would lead to a more uniform distribution of states over the energy landscape, and there might be a sweet spot for the temperature that perfectly balances stability of the attractors with occasional barrier crossings, just as in the RR phenomenon. Moreover, during the equilibration process, it might be advantageous to start with a high temperature (noise level) and then to slowly decrease it, as in Simulated Annealing \cite{van1987simulated,bertsimas1993simulated}.

Note, however, that in our SBM model the system's dynamics cannot be visualized as a simple probabilistic downhill relaxation within the energy landscape $E(\mathbf{s})$, because we are using a synchronous update of all neurons and non-symmetric weights. For example, in NRooks systems, each n-cycle attractor can have another energy, but the energy is the same for all states that belong to the same attractor. Despite of this 'energetical degeneration' of states within a given n-cycle, the system is not randomly jumping between those states, as it would be expected from a thermal system, but it is running through the state sequence in a perfectly deterministic way.

\NI In this work, we have mainly focused on probabilistic SBMs as model systems of recurrent neural networks. However, we have shown that for sufficiently large weight magnitudes $w$ the neurons operate in the saturation regime of the logistic activation function and therefore the SBMs behave quasi-deterministic. In this regime, the RR curves of SBMs turned out to be extremely similar to those of networks with the same weight matrix, but with deterministic tanh-neurons. Also, we have demonstrated that attractor switching by noise pulses works equally well for tanh-networks with continuous outputs. Nevertheless, regarding the details of the RR phenomenon, we expect future work to reveal some fundamental differences between probabilistic and deterministic RNNs. In particular, while the SBM neurons turn into independent random generators when the general weight magnitude $w$ of their connections is turned down, deterministic networks can - even for small $w$ - produce complex dynamical attractor states. It is not clear at present how sensitive those attractors react on externally injected noise.

\NI Neural networks, both artificial and biological, have a tendency to become trapped in low-entropy dynamical attractors, which correspond to repetitive, predictable patterns of activity \cite{khona2022attractor}. These attractors are often associated with stable cognitive states or established perceptual interpretations \cite{beer2024revealing}. In particular, it has been shown that during spontaneous activity the brain does not randomly change between all theoretically possible states, but rather samples from the realm of possible sensory responses \cite{luczak2009spontaneous,schilling2024deep}. This indicates that the brain's spontaneous activity encompasses a spectrum of potential responses to stimuli, effectively preconfiguring the neural landscape for incoming sensory information. While stability and predictability are crucial for efficient functioning and reliable behavior, they can also limit the flexibility and adaptability of the system \cite{khona2022attractor,beer2024revealing}. This is particularly problematic in contexts requiring learning, creativity, and the generation of novel ideas \cite{sandamirskaya2014dynamic}.

\NI The introduction of noise into neural networks has been suggested as a mechanism to overcome the limitations imposed by these low-entropy attractors \cite{hinton1993keeping}. Noise, in this context, refers to stochastic fluctuations that perturb the network's activity, pushing it out of stable attractor states and into new regions of the state space \cite{bishop1995training}. This process can enhance the network's ability to explore a wider range of potential states \cite{sietsma1991creating}, thereby increasing its entropy and promoting the discovery of novel solutions or interpretations.

\NI Biological neural systems, such as the human brain, provide compelling evidence for the utility of noise in cognitive processes. The brain is inherently noisy, with intrinsic fluctuations occurring at multiple levels, from ion channel gating to synaptic transmission and neural firing \cite{faisal2008noise}. This noise is not merely a byproduct of biological imperfection; rather, it plays a functional role in various cognitive tasks \cite{mcdonnell2011benefits}. For example, noise-induced variability in neural firing can enhance sensory perception by enabling the brain to detect weak signals that would otherwise be drowned out by deterministic activity \cite{deco2009stochastic}. Similarly, stochastic resonance, a phenomenon where noise enhances the response of a system to weak inputs, has been observed in various sensory systems \cite{moss2004stochastic,stein2005neuronal,ward2013thalamus}.

\NI Noise also facilitates learning and plasticity. During development, random fluctuations in neural activity contribute to the refinement of neural circuits, allowing for the fine-tuning of synaptic connections based on experience \cite{marzola2023exploring,zhang2021noise,fang2020noise}. In adulthood, noise can help the brain escape from local minima during learning processes, thereby preventing overfitting to specific patterns and promoting generalization \cite{zhang2021noise,fang2020noise}. This is particularly relevant in the context of reinforcement learning, where exploration of the state space is crucial for finding optimal strategies \cite{weng2020exploration,hao2023exploration}.

\NI Moreover, noise-induced transitions between attractor states can support cognitive flexibility and creativity. For instance, the ability to switch between different interpretations of ambiguous stimuli \cite{panagiotaropoulos2013common}, or to generate novel ideas, relies on the brain's capacity to break free from dominant attractor states and explore alternative possibilities \cite{wu2020charting,jaimes2022multistability}. This is consistent with the observation that certain cognitive disorders, characterized by rigidity and a lack of flexibility (e.g., autism, obsessive-compulsive disorder), are associated with reduced neural noise and hyper-stable attractor dynamics \cite{dwyer2024neural,watanabe2019neuroanatomical}.

\NI In conclusion, the investigations detailed in our study firmly establish Recurrence Resonance (RR) as a genuine emergent phenomenon within neural dynamics: the mutual information of the system is increased by the addition of noise that itself has zero mutual information. Hence, the application of optimal noise levels can transform neural systems from states of minimal information processing capabilities to significantly enhanced states where information flow is not only possible but also maximized. This effect, whereby noise beneficially modifies system dynamics, underscores the complex and non-intuitive nature of neural information processing, presenting noise not merely as a disruptor but as a critical facilitator of dynamic neural activity. This finding opens up new avenues for exploiting noise in the design and enhancement of neural network models, particularly in areas demanding robust and adaptive information processing.

\NI The introduction of noise into neural networks can be seen as a fundamental mechanism by which the brain enhances its cognitive capabilities. By destabilizing low-entropy attractors and promoting the exploration of new states, noise enables learning, perception, and creativity. This perspective not only aligns with empirical findings from neuroscience but also offers a theoretical framework for understanding how complex cognitive functions can emerge from the interplay between deterministic and stochastic processes in neural systems.

\NI Furthermore, the insights gained from our study provide a valuable foundation for advancing artificial intelligence (AI) technologies, particularly in the realms of reservoir computing and machine learning. Reservoir computing, which leverages the dynamic behavior of recurrent neural networks, can benefit from the strategic introduction of noise to enhance its computational power and adaptability. Similarly, machine learning models can incorporate noise to avoid overfitting, explore diverse solution spaces, and improve generalization. By integrating these principles, AI systems can emulate the brain’s ability to learn and adapt in complex, unpredictable environments, leading to more robust and innovative technological solutions. This convergence of neuroscience and AI not only deepens our understanding of cognitive processes but may also drive the development of next-generation intelligent systems capable of solving real-world problems with unprecedented efficiency and creativity.

\section{Additional Information}

\subsection{Author contributions}

CM conceived the study, implemented the methods, evaluated the data, and wrote the paper. PK conceived the study, discussed the results, acquired funding and wrote the paper. AS discussed the results and acquired funding. AM discussed the results and provided resources.

\subsection{Funding}
This work was funded by the Deutsche Forschungsgemeinschaft (DFG, German Research Foundation): KR\,5148/3-1 (project number 510395418), KR\,5148/5-1 (project number 542747151), and GRK\,2839 (project number 468527017) to PK, and grant SCHI\,1482/3-1 (project number 451810794) to AS.

\subsection{Competing interests statement}
The authors declare no competing interests.

\subsection{Data availability statement}
The complete data and analysis programs will be made available upon reasonable request.

\subsection{Third party rights}
All material used in the paper are the intellectual property of the authors.


\bibliographystyle{unsrt}

\begin{thebibliography}{10}

\bibitem{krauss2019recurrence}
Patrick Krauss, Karin Prebeck, Achim Schilling, and Claus Metzner.
\newblock Recurrence resonance” in three-neuron motifs.
\newblock {\em Frontiers in computational neuroscience}, 13, 2019.

\bibitem{lecun2015deep}
Yann LeCun, Yoshua Bengio, and Geoffrey Hinton.
\newblock Deep learning.
\newblock {\em nature}, 521(7553):436--444, 2015.

\bibitem{alzubaidi2021review}
Laith Alzubaidi, Jinglan Zhang, Amjad~J Humaidi, Ayad Al-Dujaili, Ye~Duan, Omran Al-Shamma, Jos{\'e} Santamar{\'\i}a, Mohammed~A Fadhel, Muthana Al-Amidie, and Laith Farhan.
\newblock Review of deep learning: Concepts, cnn architectures, challenges, applications, future directions.
\newblock {\em Journal of big Data}, 8(1):1--74, 2021.

\bibitem{maheswaranathan2019universality}
Niru Maheswaranathan, Alex~H Williams, Matthew~D Golub, Surya Ganguli, and David Sussillo.
\newblock Universality and individuality in neural dynamics across large populations of recurrent networks.
\newblock {\em Advances in neural information processing systems}, 2019:15629, 2019.

\bibitem{schafer2006recurrent}
Anton~Maximilian Sch{\"a}fer and Hans~Georg Zimmermann.
\newblock Recurrent neural networks are universal approximators.
\newblock In {\em International Conference on Artificial Neural Networks}, pages 632--640. Springer, 2006.

\bibitem{jaeger2001echo}
Herbert Jaeger.
\newblock The “echo state” approach to analysing and training recurrent neural networks-with an erratum note.
\newblock {\em Bonn, Germany: German National Research Center for Information Technology GMD Technical Report}, 148(34):13, 2001.

\bibitem{schuecker2018optimal}
Jannis Schuecker, Sven Goedeke, and Moritz Helias.
\newblock Optimal sequence memory in driven random networks.
\newblock {\em Physical Review X}, 8(4):041029, 2018.

\bibitem{busing2010connectivity}
Lars B{\"u}sing, Benjamin Schrauwen, and Robert Legenstein.
\newblock Connectivity, dynamics, and memory in reservoir computing with binary and analog neurons.
\newblock {\em Neural computation}, 22(5):1272--1311, 2010.

\bibitem{dambre2012information}
Joni Dambre, David Verstraeten, Benjamin Schrauwen, and Serge Massar.
\newblock Information processing capacity of dynamical systems.
\newblock {\em Scientific reports}, 2(1):1--7, 2012.

\bibitem{wallace2013randomly}
Edward Wallace, Hamid~Reza Maei, and Peter~E Latham.
\newblock Randomly connected networks have short temporal memory.
\newblock {\em Neural computation}, 25(6):1408--1439, 2013.

\bibitem{gonon2021fading}
Lukas Gonon and Juan-Pablo Ortega.
\newblock Fading memory echo state networks are universal.
\newblock {\em Neural Networks}, 138:10--13, 2021.

\bibitem{farrell2022gradient}
Matthew Farrell, Stefano Recanatesi, Timothy Moore, Guillaume Lajoie, and Eric Shea-Brown.
\newblock Gradient-based learning drives robust representations in recurrent neural networks by balancing compression and expansion.
\newblock {\em Nature Machine Intelligence}, 4(6):564--573, 2022.

\bibitem{kadmon2015transition}
Jonathan Kadmon and Haim Sompolinsky.
\newblock Transition to chaos in random neuronal networks.
\newblock {\em Physical Review X}, 5(4):041030, 2015.

\bibitem{wang2011fisher}
X~Rosalind Wang, Joseph~T Lizier, and Mikhail Prokopenko.
\newblock Fisher information at the edge of chaos in random boolean networks.
\newblock {\em Artificial life}, 17(4):315--329, 2011.

\bibitem{boedecker2012information}
Joschka Boedecker, Oliver Obst, Joseph~T Lizier, N~Michael Mayer, and Minoru Asada.
\newblock Information processing in echo state networks at the edge of chaos.
\newblock {\em Theory in Biosciences}, 131(3):205--213, 2012.

\bibitem{langton1990computation}
Chris~G Langton.
\newblock Computation at the edge of chaos: Phase transitions and emergent computation.
\newblock {\em Physica D: Nonlinear Phenomena}, 42(1-3):12--37, 1990.

\bibitem{natschlager2005edge}
Thomas Natschl{\"a}ger, Nils Bertschinger, and Robert Legenstein.
\newblock At the edge of chaos: Real-time computations and self-organized criticality in recurrent neural networks.
\newblock {\em Advances in neural information processing systems}, 17:145--152, 2005.

\bibitem{legenstein2007edge}
Robert Legenstein and Wolfgang Maass.
\newblock Edge of chaos and prediction of computational performance for neural circuit models.
\newblock {\em Neural networks}, 20(3):323--334, 2007.

\bibitem{bertschinger2004real}
Nils Bertschinger and Thomas Natschl{\"a}ger.
\newblock Real-time computation at the edge of chaos in recurrent neural networks.
\newblock {\em Neural computation}, 16(7):1413--1436, 2004.

\bibitem{schrauwen2009computational}
Benjamin Schrauwen, Lars Buesing, and Robert Legenstein.
\newblock On computational power and the order-chaos phase transition in reservoir computing.
\newblock In {\em 22nd Annual conference on Neural Information Processing Systems (NIPS 2008)}, volume~21, pages 1425--1432. NIPS Foundation, 2009.

\bibitem{toyoizumi2011beyond}
Taro Toyoizumi and LF~Abbott.
\newblock Beyond the edge of chaos: Amplification and temporal integration by recurrent networks in the chaotic regime.
\newblock {\em Physical Review E}, 84(5):051908, 2011.

\bibitem{kaneko1994evolution}
Kunihiko Kaneko and Junji Suzuki.
\newblock Evolution to the edge of chaos in an imitation game.
\newblock In {\em Artificial life III}. Citeseer, 1994.

\bibitem{sole1995information}
Ricard~V Sol{\'e} and Octavio Miramontes.
\newblock Information at the edge of chaos in fluid neural networks.
\newblock {\em Physica D: Nonlinear Phenomena}, 80(1-2):171--180, 1995.

\bibitem{haruna2019optimal}
Taichi Haruna and Kohei Nakajima.
\newblock Optimal short-term memory before the edge of chaos in driven random recurrent networks.
\newblock {\em Physical Review E}, 100(6):062312, 2019.

\bibitem{ichikawa2021short}
Kohei Ichikawa and Kunihiko Kaneko.
\newblock Short-term memory by transient oscillatory dynamics in recurrent neural networks.
\newblock {\em Physical Review Research}, 3(3):033193, 2021.

\bibitem{rajan2010stimulus}
Kanaka Rajan, LF~Abbott, and Haim Sompolinsky.
\newblock Stimulus-dependent suppression of chaos in recurrent neural networks.
\newblock {\em Physical Review E}, 82(1):011903, 2010.

\bibitem{jaeger2014controlling}
Herbert Jaeger.
\newblock Controlling recurrent neural networks by conceptors.
\newblock {\em arXiv preprint arXiv:1403.3369}, 2014.

\bibitem{haviv2019understanding}
Doron Haviv, Alexander Rivkind, and Omri Barak.
\newblock Understanding and controlling memory in recurrent neural networks.
\newblock In {\em International Conference on Machine Learning}, pages 2663--2671. PMLR, 2019.

\bibitem{molgedey1992suppressing}
Lutz Molgedey, J~Schuchhardt, and Heinz~G Schuster.
\newblock Suppressing chaos in neural networks by noise.
\newblock {\em Physical review letters}, 69(26):3717, 1992.

\bibitem{ikemoto2018noise}
Shuhei Ikemoto, Fabio DallaLibera, and Koh Hosoda.
\newblock Noise-modulated neural networks as an application of stochastic resonance.
\newblock {\em Neurocomputing}, 277:29--37, 2018.

\bibitem{bonsel2021control}
Florian B{\"o}nsel, Patrick Krauss, Claus Metzner, and Marius~E Yamakou.
\newblock Control of noise-induced coherent oscillations in time-delayed neural motifs.
\newblock {\em arXiv preprint arXiv:2106.11361}, 2021.

\bibitem{metzner2022dynamics}
Claus Metzner and Patrick Krauss.
\newblock Dynamics and information import in recurrent neural networks.
\newblock {\em Frontiers in Computational Neuroscience}, 16, 2022.

\bibitem{barak2017recurrent}
Omri Barak.
\newblock Recurrent neural networks as versatile tools of neuroscience research.
\newblock {\em Current opinion in neurobiology}, 46:1--6, 2017.

\bibitem{song2005highly}
Sen Song, Per~Jesper Sj{\"o}str{\"o}m, Markus Reigl, Sacha Nelson, and Dmitri~B Chklovskii.
\newblock Highly nonrandom features of synaptic connectivity in local cortical circuits.
\newblock {\em PLoS biology}, 3(3):e68, 2005.

\bibitem{brunel2016cortical}
Nicolas Brunel.
\newblock Is cortical connectivity optimized for storing information?
\newblock {\em Nature neuroscience}, 19(5):749--755, 2016.

\bibitem{narang2017exploring}
Sharan Narang, Erich Elsen, Gregory Diamos, and Shubho Sengupta.
\newblock Exploring sparsity in recurrent neural networks.
\newblock {\em arXiv preprint arXiv:1704.05119}, 2017.

\bibitem{gerum2020sparsity}
Richard~C Gerum, Andr{\'e} Erpenbeck, Patrick Krauss, and Achim Schilling.
\newblock Sparsity through evolutionary pruning prevents neuronal networks from overfitting.
\newblock {\em Neural Networks}, 128:305--312, 2020.

\bibitem{folli2018effect}
Viola Folli, Giorgio Gosti, Marco Leonetti, and Giancarlo Ruocco.
\newblock Effect of dilution in asymmetric recurrent neural networks.
\newblock {\em Neural Networks}, 104:50--59, 2018.

\bibitem{krauss2019analysis}
Patrick Krauss, Alexandra Zankl, Achim Schilling, Holger Schulze, and Claus Metzner.
\newblock Analysis of structure and dynamics in three-neuron motifs.
\newblock {\em Frontiers in Computational Neuroscience}, 13:5, 2019.

\bibitem{krauss2019weight}
Patrick Krauss, Marc Schuster, Verena Dietrich, Achim Schilling, Holger Schulze, and Claus Metzner.
\newblock Weight statistics controls dynamics in recurrent neural networks.
\newblock {\em PloS one}, 14(4):e0214541, 2019.

\bibitem{schilling2022intrinsic}
Achim Schilling, Richard Gerum, Claus Metzner, Andreas Maier, and Patrick Krauss.
\newblock Intrinsic noise improves speech recognition in a computational model of the auditory pathway.
\newblock {\em Frontiers in Neuroscience}, 16:908330, 2022.

\bibitem{krauss2016stochastic}
Patrick Krauss, Konstantin Tziridis, Claus Metzner, Achim Schilling, Ulrich Hoppe, and Holger Schulze.
\newblock Stochastic resonance controlled upregulation of internal noise after hearing loss as a putative cause of tinnitus-related neuronal hyperactivity.
\newblock {\em Frontiers in neuroscience}, 10:597, 2016.

\bibitem{schilling2021stochastic}
Achim Schilling, Konstantin Tziridis, Holger Schulze, and Patrick Krauss.
\newblock The stochastic resonance model of auditory perception: A unified explanation of tinnitus development, zwicker tone illusion, and residual inhibition.
\newblock {\em Progress in Brain Research}, 262:139--157, 2021.

\bibitem{schilling2023predictive}
Achim Schilling, William Sedley, Richard Gerum, Claus Metzner, Konstantin Tziridis, Andreas Maier, Holger Schulze, Fan-Gang Zeng, Karl~J Friston, and Patrick Krauss.
\newblock Predictive coding and stochastic resonance as fundamental principles of auditory phantom perception.
\newblock {\em Brain}, 146(12):4809--4825, 2023.

\bibitem{metzner2024quantifying}
Claus Metzner, Marius~E Yamakou, Dennis Voelkl, Achim Schilling, and Patrick Krauss.
\newblock Quantifying and maximizing the information flux in recurrent neural networks.
\newblock {\em Neural Computation}, 36(3):351--384, 2024.

\bibitem{katz1972coverage}
Leo Katz and Milton Sobel.
\newblock Coverage of generalized chess boards by randomly placed rooks.
\newblock In {\em Proceedings of the Sixth Berkeley Symposium on Mathematical Statistics and Probability, Volume 3: Probability Theory}, volume~6, pages 555--565. University of California Press, 1972.

\bibitem{hopfield1982neural}
John~J Hopfield.
\newblock Neural networks and physical systems with emergent collective computational abilities.
\newblock {\em Proceedings of the national academy of sciences}, 79(8):2554--2558, 1982.

\bibitem{zwanzig1992levinthal}
Robert Zwanzig, Attila Szabo, and Biman Bagchi.
\newblock Levinthal's paradox.
\newblock {\em Proceedings of the National Academy of Sciences}, 89(1):20--22, 1992.

\bibitem{karplus1997levinthal}
Martin Karplus.
\newblock The levinthal paradox: yesterday and today.
\newblock {\em Folding and design}, 2:S69--S75, 1997.

\bibitem{honig1999protein}
Barry Honig.
\newblock Protein folding: from the levinthal paradox to structure prediction.
\newblock {\em Journal of Molecular Biology}, 293(2):283--293, 1999.

\bibitem{bryngelson1995funnels}
Joseph~D Bryngelson, Jos{\'e}~Nelson Onuchic, Nicholas~D Socci, and Peter~G Wolynes.
\newblock Funnels, pathways, and the energy landscape of protein folding: a synthesis.
\newblock {\em Proteins: Structure, Function, and Bioinformatics}, 21(3):167--195, 1995.

\bibitem{martinez2014introducing}
Leandro Mart{\'\i}nez.
\newblock Introducing the levinthal’s protein folding paradox and its solution.
\newblock {\em Journal of Chemical Education}, 91(11):1918--1923, 2014.

\bibitem{wolynes2015evolution}
Peter~G Wolynes.
\newblock Evolution, energy landscapes and the paradoxes of protein folding.
\newblock {\em Biochimie}, 119:218--230, 2015.

\bibitem{roder2019energy}
Konstantin R{\"o}der, Jerelle~A Joseph, Brooke~E Husic, and David~J Wales.
\newblock Energy landscapes for proteins: From single funnels to multifunctional systems.
\newblock {\em Advanced Theory and Simulations}, 2(4):1800175, 2019.

\bibitem{hooper2000central}
Scott~L Hooper.
\newblock Central pattern generators.
\newblock {\em Current Biology}, 10(5):R176--R179, 2000.

\bibitem{marder2001central}
Eve Marder and Dirk Bucher.
\newblock Central pattern generators and the control of rhythmic movements.
\newblock {\em Current biology}, 11(23):R986--R996, 2001.

\bibitem{van1987simulated}
Peter~JM Van~Laarhoven, Emile~HL Aarts, Peter~JM van Laarhoven, and Emile~HL Aarts.
\newblock {\em Simulated annealing}.
\newblock Springer, 1987.

\bibitem{bertsimas1993simulated}
Dimitris Bertsimas and John Tsitsiklis.
\newblock Simulated annealing.
\newblock {\em Statistical science}, 8(1):10--15, 1993.

\bibitem{khona2022attractor}
Mikail Khona and Ila~R Fiete.
\newblock Attractor and integrator networks in the brain.
\newblock {\em Nature Reviews Neuroscience}, 23(12):744--766, 2022.

\bibitem{beer2024revealing}
Chen Beer and Omri Barak.
\newblock Revealing and reshaping attractor dynamics in large networks of cortical neurons.
\newblock {\em PLOS Computational Biology}, 20(1):e1011784, 2024.

\bibitem{luczak2009spontaneous}
Artur Luczak, Peter Barth{\'o}, and Kenneth~D Harris.
\newblock Spontaneous events outline the realm of possible sensory responses in neocortical populations.
\newblock {\em Neuron}, 62(3):413--425, 2009.

\bibitem{schilling2024deep}
Achim Schilling, Richard Gerum, Claudia Boehm, Jwan Rasheed, Claus Metzner, Andreas Maier, Caroline Reindl, Hajo Hamer, and Patrick Krauss.
\newblock Deep learning based decoding of single local field potential events.
\newblock {\em NeuroImage}, page 120696, 2024.

\bibitem{sandamirskaya2014dynamic}
Yulia Sandamirskaya.
\newblock Dynamic neural fields as a step toward cognitive neuromorphic architectures.
\newblock {\em Frontiers in neuroscience}, 7:276, 2014.

\bibitem{hinton1993keeping}
Geoffrey~E Hinton and Drew Van~Camp.
\newblock Keeping the neural networks simple by minimizing the description length of the weights.
\newblock In {\em Proceedings of the sixth annual conference on Computational learning theory}, pages 5--13, 1993.

\bibitem{bishop1995training}
Chris~M Bishop.
\newblock Training with noise is equivalent to tikhonov regularization.
\newblock {\em Neural computation}, 7(1):108--116, 1995.

\bibitem{sietsma1991creating}
Jocelyn Sietsma and Robert~JF Dow.
\newblock Creating artificial neural networks that generalize.
\newblock {\em Neural networks}, 4(1):67--79, 1991.

\bibitem{faisal2008noise}
A~Aldo Faisal, Luc~PJ Selen, and Daniel~M Wolpert.
\newblock Noise in the nervous system.
\newblock {\em Nature reviews neuroscience}, 9(4):292--303, 2008.

\bibitem{mcdonnell2011benefits}
Mark~D McDonnell and Lawrence~M Ward.
\newblock The benefits of noise in neural systems: bridging theory and experiment.
\newblock {\em Nature Reviews Neuroscience}, 12(7):415--425, 2011.

\bibitem{deco2009stochastic}
Gustavo Deco, Edmund~T Rolls, and Ranulfo Romo.
\newblock Stochastic dynamics as a principle of brain function.
\newblock {\em Progress in neurobiology}, 88(1):1--16, 2009.

\bibitem{moss2004stochastic}
Frank Moss, Lawrence~M Ward, and Walter~G Sannita.
\newblock Stochastic resonance and sensory information processing: a tutorial and review of application.
\newblock {\em Clinical neurophysiology}, 115(2):267--281, 2004.

\bibitem{stein2005neuronal}
Richard~B Stein, E~Roderich Gossen, and Kelvin~E Jones.
\newblock Neuronal variability: noise or part of the signal?
\newblock {\em Nature Reviews Neuroscience}, 6(5):389--397, 2005.

\bibitem{ward2013thalamus}
Lawrence~M Ward.
\newblock The thalamus: gateway to the mind.
\newblock {\em Wiley Interdisciplinary Reviews: Cognitive Science}, 4(6):609--622, 2013.

\bibitem{marzola2023exploring}
Patr{\'\i}cia Marzola, Thayza Melzer, Eloisa Pavesi, Joana Gil-Mohapel, and Patricia~S Brocardo.
\newblock Exploring the role of neuroplasticity in development, aging, and neurodegeneration.
\newblock {\em Brain Sciences}, 13(12):1610, 2023.

\bibitem{zhang2021noise}
Chi Zhang, Danke Zhang, and Armen Stepanyants.
\newblock Noise in neurons and synapses enables reliable associative memory storage in local cortical circuits.
\newblock {\em Eneuro}, 8(1), 2021.

\bibitem{fang2020noise}
Ying Fang, Zhaofei Yu, and Feng Chen.
\newblock Noise helps optimization escape from saddle points in the synaptic plasticity.
\newblock {\em Frontiers in neuroscience}, 14:343, 2020.

\bibitem{weng2020exploration}
Lilian Weng.
\newblock Exploration strategies in deep reinforcement learning.
\newblock {\em lilianweng. github. io/lil-log}, 2020.

\bibitem{hao2023exploration}
Jianye Hao, Tianpei Yang, Hongyao Tang, Chenjia Bai, Jinyi Liu, Zhaopeng Meng, Peng Liu, and Zhen Wang.
\newblock Exploration in deep reinforcement learning: From single-agent to multiagent domain.
\newblock {\em IEEE Transactions on Neural Networks and Learning Systems}, 2023.

\bibitem{panagiotaropoulos2013common}
Theofanis~I Panagiotaropoulos, Vishal Kapoor, Nikos~K Logothetis, and Gustavo Deco.
\newblock A common neurodynamical mechanism could mediate externally induced and intrinsically generated transitions in visual awareness.
\newblock {\em PLoS One}, 8(1):e53833, 2013.

\bibitem{wu2020charting}
Yihan Wu and Wilma Koutstaal.
\newblock Charting the contributions of cognitive flexibility to creativity: Self-guided transitions as a process-based index of creativity-related adaptivity.
\newblock {\em PloS one}, 15(6):e0234473, 2020.

\bibitem{jaimes2022multistability}
R~Jaimes-Re{\'a}tegui, G~Huerta-Cuellar, JH~Garc{\'\i}a-L{\'o}pez, and AN~Pisarchik.
\newblock Multistability and noise-induced transitions in the model of bidirectionally coupled neurons with electrical synaptic plasticity.
\newblock {\em The European Physical Journal Special Topics}, pages 1--11, 2022.

\bibitem{dwyer2024neural}
Patrick Dwyer, Svjetlana Vukusic, Zachary~J Williams, Clifford~D Saron, and Susan~M Rivera.
\newblock “neural noise” in auditory responses in young autistic and neurotypical children.
\newblock {\em Journal of autism and developmental disorders}, 54(2):642--661, 2024.

\bibitem{watanabe2019neuroanatomical}
Takamitsu Watanabe, Rebecca~P Lawson, Ylva~SE Walld{\'e}n, and Geraint Rees.
\newblock A neuroanatomical substrate linking perceptual stability to cognitive rigidity in autism.
\newblock {\em Journal of Neuroscience}, 39(33):6540--6554, 2019.

\end{thebibliography}


\newpage
\newpage

\newpage

\newpage


\end{document}